\documentclass[final]{cvpr}

\usepackage{times}
\usepackage{epsfig}
\usepackage{graphicx}
\usepackage{amsmath}
\usepackage{amssymb}
\usepackage{algorithm}
\usepackage{booktabs}
\usepackage{microtype}
\usepackage{algpseudocode}
\usepackage{multirow}
\usepackage{float}
\usepackage[toc,page]{appendix}
\usepackage{flushend}
\usepackage{amsfonts}
\usepackage{bbm}

\usepackage{subfigure}
\usepackage{caption}

\usepackage[pagebackref=true,breaklinks=true,colorlinks,bookmarks=false]{hyperref}
\usepackage{cleveref}

\title{Re-purposing Perceptual Hashing based Client Side Scanning for Physical Surveillance}
  
\author{
    Ashish Hooda\\
    University of Wisconsin\\
    Madison\\
    ahooda@wisc.edu
  \and
    Andrey Labunets\\
    University of California\\
    San Diego\\
    alabunets@ucsd.edu
    \and
    Tadayoshi Kohno\\
    University of Washington\\
    Seattle\\
    yoshi@cs.washington.edu
    \and
    Earlence Fernandes\\
    University of California\\
    San Diego\\
    efernandes@ucsd.edu
}

\usepackage{fancyhdr}
\begin{document}

\maketitle
\thispagestyle{fancy}

\begin{abstract}\label{sec:abs}
Content scanning systems employ perceptual hashing algorithms to scan user content for illegal material, such as child pornography or terrorist recruitment flyers. Perceptual hashing algorithms help determine whether two images are visually similar while preserving the privacy of the input images. Several efforts from industry and academia propose to conduct content scanning on client devices such as smartphones due to the impending roll out of end-to-end encryption that will make server-side content scanning difficult. However, these proposals have met with strong criticism because of the potential for the technology to be misused and re-purposed. Our work informs this conversation by experimentally characterizing the potential for one type of misuse --- attackers manipulating the content scanning system to perform physical surveillance on target locations. Our contributions are threefold: (1) we offer a definition of physical surveillance in the context of client-side image scanning systems; (2) we experimentally characterize this risk and create a surveillance algorithm that achieves physical surveillance rates of $>40\%$ by poisoning 5\% of the perceptual hash database; (3) we experimentally study the trade-off between the robustness of client-side image scanning systems and surveillance, showing that more robust detection of illegal material leads to increased potential for physical surveillance.
\end{abstract}

\maketitle


\section{Introduction}\label{sec:intro}

Many file sharing and communication service providers scan user image data against lists of known illegal images. This helps detect child sexual abuse material (CSAM), non-consensual pornography and terrorist recruitment material~\cite{facebookscan,AppleCSAM}. The service provider will ban associated accounts and optionally report the user identities to law enforcement for further legal action, depending on the severity of the violation. Various government and non-profit bodies curate the lists of illegal content, such as the National Center for Missing and Exploited Children (NCMEC) in the US, and Internet Watch Foundation in the UK.

With the impending roll out of End-to-End encryption of user data on services such as iCloud~\cite{icloud-encryption}, this type of server-side scanning is no longer possible. Motivated by this, recent proposals instead perform scanning locally on the user's device (e.g., smartphone) before content is encrypted. Termed Client-Side Image Scanning (CSIS), these proposals employ perceptual hashing algorithms (e.g., PhotoDNA~\cite{microscan}, NeuralHash~\cite{AppleCSAM}, PDQ~\cite{facebookscan}) that convert visually similar images to similar hashes.  The client-side content scanning systems match perceptual hashes of user images against a curated database of illegal image hashes.

Perceptual hashes are not cryptographic hashes; they preserve the visual similarity of differing images --- hashing visually similar images will produce hash values that are close to each other according to a distance metric (e.g., euclidean distance). Therefore, a perceptual hash does not change (or changes only by a little amount) when the underlying images undergo transformations like re-coding or re-sizing.  



Despite the potential for this technology to curb the distribution of illegal content, there is potential for misuse. Critics of client-side image scanning have pointed out that nation states or otherwise malicious law enforcement agencies can re-purpose CSIS systems to perform surveillance and censorship on any type of content~\cite{bugsinpocket}. For example, a nation state could attempt to monitor the private content of whistleblowers and journalists. Consequently, a growing line of work in the community is exploring how CSIS systems employing perceptual hashing like Apple's NeuralHash or Facebook's PDQ could be misused~\cite{jain2022adversarial,cryptoeprint:2021/1531,bugsinpocket,breakNeuralHash}.


We contribute to this line of work by introducing the notion of physical surveillance in CSIS and then experimentally characterizing the physical surveillance risks. Specifically, we define physical surveillance as the capability of an attacker to visually monitor a target scene as if the attacker has placed a camera at that location. We empirically characterize the extent to which the attacker's capability approximates this idea. Our primary insight is that the photographs that users take of a scene can become accessible to an attacker who manipulates the content scanning system database. 

Consider a scene in an art gallery that the attacker wishes to monitor. Gallery visitors will take photographs of themselves and of artwork. Due to the presence of CSIS on their phones, each image will be matched against a database of illicit image hashes. If the attacker is able to strategically manipulate this database of illicit hashes, then they can induce matches that would result in the user's images being transmitted to the service provider and getting decrypted. With the raw images, the attacker gains a capability similar to them having placed a camera at that physical location. The caveat is that this surveillance capability only approximates a real camera because the images they get access to depend on the photographs that users take and the inexact matching process. Our work experimentally characterizes the extent to which such a physical surveillance attack is possible. 

At this point, the attacker can perform additional analyses on the images to extract different types of information. For example, they could run the image through a face recognition system~\cite{facerecnyt,facerecwallstreet,wenger2021sok}; or if the CSIS system also reports user identity, they can automatically place a specific user at a specific physical location.


The attacker's high-level strategy is to poison the curated hash database of illegal content with hashes that are crafted to match the images that users might take at target physical locations. This requires overcoming a few challenges. First, perceptual hashing-based CSIS is robust to only small input image modifications. Physical surveillance requires detection of a wide variety of images of a scene. Second, the attacker's poison hashes have to correspond to illegal content. Otherwise, a human curator of the database can easily flag images submitted by the attacker as being irrelevant. 

We address these challenges in the following ways. First, we contribute an algorithm that computes the optimal set of poison hashes using a k-modes clustering process on the target physical scene hashes. Second, we rely on the observation that perceptual hashing algorithms are susceptible to adversarial examples~\cite{breakNeuralHash,jain2022adversarial,cryptoeprint:2021/1531,dolhansky2020adversarial}. It is possible to craft two images that are visually different but their hashes are very similar. Specifically, starting from a known illegal image, we modify it so that it still looks like illegal content to a human curator but its corresponding hash will collide with the hashes of a specific physical scene.

A final challenge is the experimental setup and datasets required for such an analysis. We do not experiment with illegal images, but rather, use existing datasets from the ML community to approximate the concept of illegal content. Specifically, we use `superclasses' from the Imagenet dataset to catergorize illegal and benign images~\cite{tsipras2020imagenet}. We also collect images of multiple target physical locations via scraping user images uploaded on Instagram and also manually capturing photographs~\cite{gomez2018learning}.

We focus our analysis on Facebook PDQ~\cite{facebookscan} and Apple NeuralHash~\cite{AppleCSAM}. PDQ is a perceptual hash function that is currently deployed as a server-side scanning system, but it is useful in a client-side deployment scenario as well. Apple's NeuralHash system is the first instance of a commercial client-side image scanning system with the potential to affect millions of users. At the time of release, the system drew strong criticism from various groups~\cite{bugsinpocket}, leading to Apple pausing the rollout of the system~\cite{pausecsam}. Much of this criticism regarding surveillance was hypothetical. Our work provides experimental evidence of the extent to which attackers could be successful with physical surveillance goals, should CSIS systems be deployed to the public at large scale.


\noindent\textbf{Contributions.}
\begin{enumerate}
    \item We define a physical surveillance threat model for perceptual hashing-based client-side image scanning systems.
    \item We introduce a poisoning attack that can covertly re-purpose CSIS systems to perform physical surveillance. Our attacks are able to achieve $>40\%$ surveillance by poisoning just $5\%$ of the illegal content database.
    \item We characterize the trade-off between the system's ability to detect the designated illegal content and the potential to perform the covert physical surveillance: more robust detection of illegal content leads to higher risks of physical surveillance.
\end{enumerate}

\noindent\textbf{Ethical Considerations.}
Our work serves to inform the conversation around the benefits and risks of client-side image scanning technologies. Specifically, we contribute a set of experimental analyses that showcase the extent to which an adversary might misuse this technology to perform physical surveillance. We do not take a concrete stance on the issue of whether CSIS technology in its current form should be widely deployed; nor do we wish to imply that because physical surveillance is possible, such technology should never be deployed. Rather, we agree that curbing illegal content like CSAM does require technological innovation, but it has to be balanced with an understanding of the inherent risks. Furthermore, we do not condone physical surveillance and do not propose these surveillance attacks with the goal of equipping attackers. Therefore, we will not be releasing attack code publicly but will manage requests on a case-by-case basis for sharing the code in the interest of scientific exploration and discovery.

\section{Background}\label{sec:back}

\subsection{Client-Side Image Scanning Overview}\label{subsec:csis}

Client-side image scanning (CSIS) systems are part of the public debate around whether law enforcement should have special access to plaintext communications. A recent shift towards end-to-end encrypted communications has created barriers for law enforcement and service providers to detect illicit content, such as child exploitation and terrorism imagery. Client-side Image scanning is a set of techniques to selectively relax end-to-end encryption guarantees depending on the content being transmitted in an encrypted channel. By scanning content on the client device before it is encrypted, the technology provides a way for service providers and law enforcement to open up encryptions if the underlying content matches known illegal content. At this point, the content provider can report the user's identity to law enforcement for further investigation.


\subsection{Client-Side Image Scanning and Perceptual Hashing Formal Definition}\label{subsec:cssph}
A naïve approach to find examples of illicit images on a device would be to search for exact matches of those examples using a hash function, such as a cryptographic hash function $\mathcal{H} : \mathcal{X} \rightarrow \{0,1\}^n$, where $\mathcal{X}$ is raw image data and $n$ is a hash size. However, such a system would only find exact matches of the content and would be trivially bypassed due to collision resistance of the cryptographic hash functions: small changes in the input $\mathcal{X}$ will result in a different value $\mathcal{H}$. Instead, current systems use a different type of hashing, a perceptual hashing, with some degree of invariance to match examples that look visually similar to humans. We formally define the CSIS and its properties below, similar to the existing works on perceptual hashing \cite{cryptoeprint:2021/1531,breakNeuralHash,jain2022adversarial}.

For the raw image $\mathcal{X}$ and a length $n$, a perceptual hashing function is defined as $\mathcal{P} : \mathcal{X} \rightarrow \{0,1\}^n$. Here the length of the bit string depends on the type of perceptual hashing algorithm. Apple's Neural Hash computes hashes that are bit strings of length $96$~\cite{AppleCSAM}. Similarity between perceptual hashes is represented by a distance metric, $\mathcal{D} : \mathcal{P} \times \mathcal{P} \rightarrow \mathbb{R}$, which is typically hamming distance. A CSIS system $\mathcal{S}$ is defined by a Perceptual Hashing function $\mathcal{P}$, a distance metric $\mathcal{D}$, a threshold $t \in \mathbb{R}$, and a database $\mathcal{C} \subset \{0,1\}^n$ of hashes for the illicit content examples.

CSIS systems detect illicit images by matching perceptual hashes of a user's images against a database of hashes $\mathcal{C}$ computed using a curated list of illicit images (CSAM etc). When a user uploads an image $\mathcal{X}$, the CSIS systems flags it if the image's perceptual hash has a distance less than a threshold $t$ from at least one hash $c \in \mathcal{C}$ in the database, i.e.,
$\mathcal{D}(\mathcal{P}(\mathcal{X}), c) \leq t$. Apple NeuralHash~\cite{AppleCSAM}, Microsoft’s PhotoDNA~\cite{microscan}, and Facebook’s PDQ~\cite{facebookscan} are notable example of perceptual hashing algorithms used for image scanning.

Even though only detected images are flagged, the operational and threat landscape is much more nuanced. In most common cases, operation of CSIS involves a Service Provider deploying the protocol within some messaging application, Content Curators maintaining a database of illicit content, and Law Enforcement. CSIS would flag certain images, and they would be sent to a human for review at the service provider. This human review process is required to deal with false positives present in any perceptual hashing function. Therefore, both flagged private images and the list of illicit content hashes could be accessed by multiple parties besides just the service provider itself. In fact, Apple's August 2021 Proposal explicitly requires multiple attested organizations to vet the submitted illegal image hashes~\cite{AppleCSAM}.

\subsection{Hash Collisions}\label{subsec:hashcollison}
Unlike cryptographic hash functions, a Perceptual Hashing function $\mathcal{P}$ needs to perform fuzzy matching i.e, two semantically similar inputs should be hashed similarly. Therefore, when we say that two images collide, we imply that two images are sufficiently different to a human. 


We assume that two inputs $X_{1}, X_{2}$ are semantically different if their $L_{2}$ distance is large: $||X_{1} - X_{2}||_2 \geq \epsilon$ for some $\epsilon$. We say that a pair of semantically different $\mathcal{X}_1$, $\mathcal{X}_2$ is a \textit{collision} if $\mathcal{D}(\mathcal{P}(\mathcal{X}_1), \mathcal{P}(\mathcal{X}_2)) \leq t$.

Collisions against Perceptual Hashing functions can be found with gradient approximation methods, for example targeted second pre-image attacks on PhotoDNA and PDQ were demonstrated in ~\cite{cryptoeprint:2021/1531}. Gradient approximation is used when operations in the perceptual function are not differentiable. When perceptual function is a machine learning model, gradients are available and more efficient attacks can be used. Gradient-based methods and more advanced generative adversarial network based preimage attacks have been shown in ~\cite{breakNeuralHash}.



\subsection{Taxonomy of Content Scanning Systems risks}\label{subsec:tax}
Existing works on CSIS security~\cite{cryptoeprint:2021/1531,breakNeuralHash} demonstrate three broad categories of threats: extraction of source images, detection evasion, and falsely triggering detection. In the first case, an attacker with access to perceptual hashes is able to invert them and obtain the source images of illicit content used to generate them, which are highly sensitive and therefore not distributed publicly~\cite{athalye2018synthesizing}. It also was shown that adversary can use gradient-based or gradient-free techniques to alter the original image in order to bypass the CSIS and to distribute illegal media content. In the third case, the adversary is able to create images, whose hash matches the database examples. In this case, private user images would be falsely flagged and sent to a human for review. An adversary who compromised a service provider can perform deanonymization of users, detect transmission of censored images, and frame unwitting users for trafficking in illicit content. Collision-based attacks in the last category is our primary focus in this paper. Performing a collision attack requires either knowledge of certain hash values in a database, which are private, or ability to add more hashes into this database. Threats in the last category realized by capable or state actors can enable some form of surveillance. In our paper, we define the notion of physical surveillance and extend the hash collision attacks to perform physical-world surveillance through the content scanning systems.

\section{Physical-world Surveillance}\label{sec:physur}

We discuss a model for physical surveillance by misusing client-side image scanning technology. Concretely, the attacker wants to monitor a physical scene in a way that approximates them installing a camera at that scene. Our core insight is that users with a CSIS system deployed on their smartphones will function as a ``crowd-sourced camera.'' Consider the following case study:

\noindent\textbf{Case Study.}\label{subsec:casestudy}
The attacker wants to monitor a room in an art gallery that has a specific painting. Visitors of this gallery will naturally take photographs on their smartphones. Note that we assume a CSIS system has been installed on the smartphone; often taking the form of an operating system service such as the case of Apple NeuralHash. Assuming the attacker has been able to poison the CSIS illicit image database with hashes specific to the surveillance location, the user-taken photographs will collide with those hashes, resulting in the attacker being able to access those photos. Assuming a large scale setup with many users, the stream of photographs matching the poison hashes approximates the notion of the attacker placing a camera in that target scene. 

With this basic surveillance capability, the attacker can conduct additional analyses to extract more information. First, if the CSIS system reports user identities (e.g., Apple NeuralHash), then the attacker learns that specific individuals were at that scene. Second, the attacker can utilize a face recognition system to identify individuals who were unwittingly photographed. 

\noindent\textbf{Challenges.}\label{subsec:challenge} To accomplish the above notion of physical surveillance, the attacker faces the following challenges. First, what is the optimal set of poison hash values to inject? Naively, the attacker could visit the target scene ahead-of-time and record every possible photo a user could take and then poison the database with hashes corresponding to those images. This is problematic because the attack is not stealthy --- there are infinitely many possible photographs for a given scene, requiring the attacker to inject a very large number of poison hashes. The attacker could try to reduce the number of poison hashes, but perceptual hashing systems are not robust to the image variations that are likely to be encountered when trying to do physical surveillance. For example, a user's photograph may include a range of perspective transformations that are difficult to anticipate. Indeed, recent work has shown that systems like Apple's NeuralHash are designed to match very specific CSAM images and are naturally not robust to large image variations --- the output of the perceptual hash algorithm will change by a lot for large image variations of a specific scene such as changes in lighting and layout~\cite{breakNeuralHash}. 

Second, how can the attacker actually poison the illicit image database? Note that the database curator relies on various agencies submitting illicit images to them for analysis and insertion into the database. If the attacker directly submitted images of a surveillance location, the curators are likely to reject the image as being irrelevant and it might raise suspicion on the attacker.

\noindent\textbf{Threat Model.}\label{subsec:threatmodel} With the above case study and challenges as motivation, we introduce the following physical surveillance threat model. The attacker is a nation-state, law enforcement agency or malicious insider at a service provider that wants to monitor physical locations. They have the ability to poison the illicit image hash database of a CSIS system by submitting candidate images to the illegal content curator for insertion into the database. Any submitted image needs to pass the validation done by the curator before it's hash is inserted into the database. The attacker also have access to the surveillance location and are capable of taking photographs and videos of that scene ahead-of-time. We assume that the CSIS system will decrypt images that match the illicit image database and then transfer those images to the attacker. This can happen in many ways --- the CSIS system is designed to automatically report any matches to a government agency; or an employee at the service provider will first get access to the decrypted images for a manual review. Our threat model generically models the attacker as any combination of government agency or malicious insider at the service provider. 

Users are unwitting participants in this model. They take photographs or videos with their phones at various locations with the mistaken assumption that their photographs are end-to-end encrypted and only accessible to themselves. The service provider installs a CSIS system on their devices as part of a mandatory OS upgrade (e.g., Apple NeuralHash was supposed to be deployed in this way~\cite{AppleCSAM}).

\begin{figure}
    \centering
    \includegraphics[width=\linewidth]{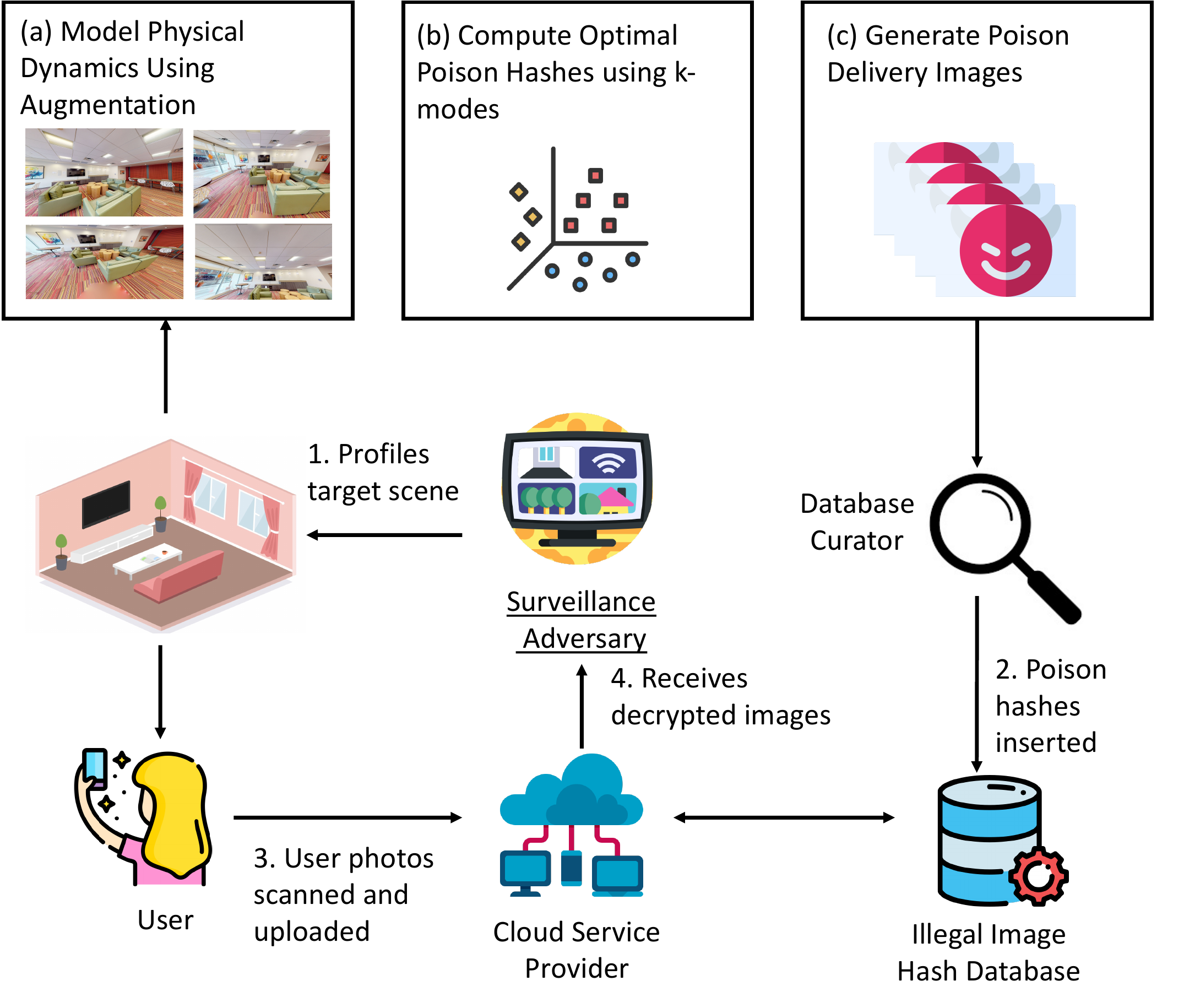}
    \caption{Pipeline of our physical surveillance attack. The attacker profiles a target scene ahead-of-time, computes a set of poison hashes using the k-modes clustering algorithm and finally inserts them into the illegal hash database using a set of crafted poison delivery images. Unwitting users take photos at the target scene that then collide with the poison hashes, resulting in the attacker gaining plaintext access to those images.}
    \label{fig:flow}
\end{figure}

\section{Design}\label{sec:design}
Our poisoning attack for physical surveillance consists of two  components corresponding to the challenges discussed above: (1) computing an optimal set of poison hashes that addresses the issue of weak perceptual hash function robustness to image transformations; (2) computing a set of images that appear like illegal content to a human curator but whose perceptual hash values are close to hashes computed in step 1 (i.e., a set of poison delivery images). We observe that step 1 of our attack framework for computing poison hashes does not depend on perceptual hash function details. The second step of computing the poison delivery images depends on whether the perceptual hash function is differentiable because the algorithm computes adversarial examples. If the hash function is not differentiable, we utilize gradient-free methods to compute the adversarial example. Figure \ref{fig:flow} describes the interacting entities and the multiple components of the attack pipeline.

\subsection{Crafting Poison Hashes}\label{subsec:craftpoison} 
Perceptual hash algorithms are invariant to small image transformations such as scaling, cropping and recoloring. They are designed to detect whether two images are syntactically similar. Note that this is different than semantic similarity which depends on high level features of the image content. This is also evident from the fact that these hash functions are not robust to semantic transformations like rotation and translation. For instance, rotating an image by only 5 degrees changes more than $10\%$ of the hash bits for both PDQ and NeuralHash ~\cite{dalins2019pdq,breakNeuralHash}. This fairly limits the ability of a CSIS system to detect transformed versions of illegal images. This is more so the case for physical surveillance which encounters high degrees of semantic transformations such as changes in perspective and viewing angles. To address this challenge, we exploit a key property of a CSIS system -- Detection of an image only depends on collisions with the best matching hash in the illegal image database. This means that regardless of the perceptual hash algorithm, a CSIS system can potentially be robust to semantic transformations as long as there is at least one hash in the illegal image database corresponding to every transformation instance. Consequently, a CSIS that uses PDQ or NeuralHash can be designed to robustly detect rotations of a specific image if we add the hashes of all possible rotations of the image to the illegal image database. However, this would not scale well to semantic transformations involved in arbitrary photographs of a scene which constitute of combinations of perspective, viewing angles and even environmental conditions. In this case, an adversary would need to poison the illegal image database with an unbounded number of hashes to account for the continuous physical transformation space. This is not feasible due to two reasons - (1) Increasing the illegal database size of a CSIS system increases the false positive rate ~\cite{jain2022adversarial}, and (2) Each hash needs to be inserted without raising suspicion from the database curators. Therefore, adding a very large number of hashes to the illegal database would make the system unusable (which will be detected by the curators) and also incur a lot of effort from the adversary. A practical physical surveillance attack needs to work with a limited number of poison insertions. Note : Inserting poisons corresponding to semantic transformations of a scene allows the CSIS system to robustly detect images from only that specific scene, and the CSIS system is only syntactically robust to all images outside that scene.

\noindent\textbf{Finding the optimal set of hashes.} As described in Section \ref{sec:back}, every CSIS system has the following four components: Perceptual Hash Function $\mathcal{P}$, illegal database $\mathcal{C}$, distance metric $\mathcal{D}$, and a distance threshold $t$. Any image $\mathcal{X}$ is flagged as illegal if there is at least one hash $c \in \mathcal{C}$ that is closer than $t$ from the image hash, i.e. $\mathcal{D}(\mathcal{P}(\mathcal{X}),c) \leq t$. Therefore, detection works as long as the distance is less than $t$. We use this property to reduce the number of poisons required to perform physical surveillance. In order to formalize this, we first need a way to represent all possible photographs of a scene. For a scene $S$, let $G_S$ be an image generating function where $I \overset{{\scriptscriptstyle\$}}{\leftarrow} G_S$  represent an image captured from the scene $S$. To successfully perform physical surveillance on scene $S$, the CSIS system should be able to detect all image generated by $G_S$. The task of generating the poisons for a CSIS using perceptual hashing : $(\mathcal{P}, \mathcal{D}, t)$ can be modelled as finding the optimal set of $k$ hashes, $\mathcal{U}_k$ such that :
\begin{equation} \label{eq:idealopt}
\begin{aligned}
  & \underset{\mathcal{U}_k}{\mathrm{arg\,max}} \:\underset{I \overset{{\scriptscriptstyle\$}}{\leftarrow} G_S}{\mathbb{E}} \left[ \mathbbm{1} \left(\min_{h \in \mathcal{U}_k} \mathcal{D}(h, \mathcal{P}(I)) < t \right) \right]
\end{aligned}
\end{equation}

Here, the inner minima finds the best matching hash for a specific image $I$ and the indicator function outputs a 1 or 0 based on whether image $I$ is detected by the set of poisons $\mathcal{U}_k$. In contrast, the outer maxima finds the best set of poisons that in expectation, maximizes the detection of images generated from $G_S$. Equation \ref{eq:idealopt} is a modified version of the covering code problem~\cite{cohen_1997}. The covering code problem finds a set of code-words in a space with a property that every other element in that space is within a fixed distance from some code-word. This problem has applications in data compression and error correction. Equation \ref{eq:idealopt} aims to find a probabilistic covering code for the distribution of images generated by $G_S$. The original covering code is a NP-complete problem ~\cite{frances1997covering}, which makes Equation \ref{eq:idealopt} even harder to solve. Therefore, we express it in a different form such that it can be solved using the k-modes clustering algorithm. We can write Equation \ref{eq:idealopt} alternatively as :
\begin{equation} \label{eq:interopt1}
\begin{aligned}
  &\underset{\mathcal{U}_k}{\mathrm{arg\,min}} \:\underset{I \overset{{\scriptscriptstyle\$}}{\leftarrow} G_S}{\mathbb{E}} \left[ \mathbbm{1} \left(\min_{h \in \mathcal{U}_k} \mathcal{D}(h, \mathcal{P}(I)) \geq t \right) \right]\\
  &\underset{\mathcal{U}_k}{\mathrm{arg\,min}} \:\underset{I \overset{{\scriptscriptstyle\$}}{\leftarrow} G_S}{\mathbb{P}}  \left(\min_{h \in \mathcal{U}_k} \mathcal{D}(h, \mathcal{P}(I)) \geq t \right)
\end{aligned}
\end{equation}

Further, since the range of the Distance metric $\mathcal{D}$ is non negative and $t\geq0$, using Markov's inequality, we get :

\begin{equation} \label{eq:marko}
\begin{aligned}
  \underset{I \overset{{\scriptscriptstyle\$}}{\leftarrow} G_S}{\mathbb{P}}  \left(\min_{h \in \mathcal{U}_k} \mathcal{D}(h, \mathcal{P}(I)) \geq t \right) \leq \frac{\underset{I \overset{{\scriptscriptstyle\$}}{\leftarrow} G_S}{\mathbb{E}}  \left[ \min_{h \in \mathcal{U}_k} \mathcal{D}(h, \mathcal{P}(I))\right]}{t}
\end{aligned}
\end{equation}

Now, since the expression in Eq. \ref{eq:interopt1} is upper bounded by Eq. \ref{eq:marko}. We can instead minimize the right side expression of Eq. \ref{eq:marko} to obtain an approximate solution for Eq. \ref{eq:interopt1}. Therefore, we optimize the following equation :
\begin{equation} \label{eq:interop2}
\begin{aligned}
  &\underset{\mathcal{U}_k}{\mathrm{arg\,min}} \:\underset{I \overset{{\scriptscriptstyle\$}}{\leftarrow} G_S}{\mathbb{E}}  \left[\min_{h \in \mathcal{U}_k} \mathcal{D}(h, \mathcal{P}(I))\right]
\end{aligned}
\end{equation}

We can approximate the expectation over the image generating function using the empirical mean. We compute the empirical mean on a large number of images captured from the scene. We can find these images by searching the internet (for a popular location) or by physically visiting the scene to capture the images. We also increase the number of images by digitally augmenting the captured images. This is done by modelling the physical transformations using perspective transforms and viewing angles. Applying this approximation using images $\{s_1, s_2, ..., s_n\}$ simplifies Equation \ref{eq:interop2} to :
\begin{equation} \label{eq:finalopt}
\begin{aligned}
  &\underset{\mathcal{U}_k}{\mathrm{arg\,min}} \:\sum\limits_{i=1}^{n} \min_{h \in \mathcal{U}_k} \mathcal{D}(h, \mathcal{P}(s_i))
\end{aligned}
\end{equation}




Equation \ref{eq:finalopt} can be solved using the $k$-modes clustering algorithm. $K$-modes is an extension of the more popular $k$-means algorithm but it works for categorical data. First, we hash each of the captured images to get a set of scene hashes $\{\mathcal{P}(s_1), \mathcal{P}(s_2),...,\mathcal{P}(s_n)\}$. Each hash in this set can be viewed as a $l$ dimension data point of binary categories, where $l$ is the length of the hash bit string (which is 96 for NeuralHash and 256 for PDQ). We now apply k-modes clustering on this set of hashes to obtain $\mathcal{U}_k$. Once we have computed the set of optimal hashes, we next look at how to insert these hashes into the illegal database.

\subsection{Computing Poison Delivery Images}

Once the optimal set of hashes is computed, they need to be inserted into the illegal content database. The illegal database is curated by organizations that are authorized to store these images. These organizations rely on various agencies including law enforcement to provide them with confiscated illegal images which can then be appended to illegal image database. Appending these illegal images to the database will then allow their detection in any subsequent scanning operations. Any illegal content curator organization that is participating in the CSIS system locally computes hashes of the entire database and provides the hashes to the service provider (since the service provider is not authorized to store the illegal images in plain text). Therefore, to insert any hash into the CSIS system, the corresponding plain text illegal image needs to be submitted to the curator. The curator can analyze the images to make sure that they match specific illegal categories before appending them into the appropriate database. Each of the optimal hashes returned by the k-modes algorithm correspond to a captured scene image (since the cluster centers of k-modes are a subset of the input data). We can submit the scene images directly to the curator for insertion into the illegal database, but this is unlikely to succeed as the scene images do not fall into any of the illegal categories and therefore, is likely to the rejected by the curator during the analysis. We get around this constraint by submitting images that perceptually look like illegal content but get hashed to a poison value --- we term these as \emph{poison delivery images}. This can be achieved by adding adversarial perturbations to a known illegal image such that it now hashes to a poison hash computed from Step 1 above. Basically, the task is to find an image $x'$ which is perceptually similar to a known illegal image $x$ i.e. $||x'-x||_2 \leq \epsilon$ for some small $\epsilon$ and $\mathcal{P}(x') \approx h$ for each $h \in \mathcal{U}_k$. Here, we borrow from existing work on generating adversarial perturbations to cause collision attacks on perceptual hashing functions~\cite{jain2022adversarial,dolhansky2020adversarial,breakNeuralHash,cryptoeprint:2021/1531}.

The CSIS system requires the used perceptual hash function to be public (since it needs to be deployed to the user device). Therefore, we can utilize gradient-based white box optimization techniques like Projected Gradient Descent to compute the delivery images. These techniques are specially effective against deep learning based perceptual hash functions such as NeuralHash. Some non deep learning based perceptual hash functions such as PDQ employ non-differentiable functions such as quantization and median. They are however, vulnerable to iterative attacks utilizing zero order gradient estimation (such as Natural Evolutionary Strategies) and also attacks involving reverse engineering the components of the hash function.

\section{Experimental Analyses of Surveillance Risks}\label{sec:eval}
We perform several experimental analyses of the extent to which an attacker can use the poisoning algorithm discussed above to achieve targeted physical surveillance. We explore surveillance success rate for various system parameters and environmental conditions.  Finally, we characterize the interplay between physical surveillance and illegal image detection and find that more robust detection of illegal material leads to better surveillance success rates for the attacker. 

\subsection{Overview}\label{subsec:quesover}
Our evaluation answers the following questions:

\begin{itemize}
    \item[\textbf{Q1.}]\label{subsec:q1} \textbf{How effectively can an adversary conduct physical surveillance for targeted locations?}\\
    We demonstrate that our poisoning attack achieves a high surveillance rate for 6 different locations relative to a false positive rate of only 1\% (i.e., the probability that a benign image's hash matches an entry in the poisoned illegal hash database). When the adversary has physical access to capture the scene images, our attack achieves a surveillance success rate of 72\% for NeuralHash and 43\% for PDQ, where we define success rate as the fraction of user photographs that get decrypted due to collisions with the poison hashes.
    
    
    \item[\textbf{Q2.}]\label{subsec:q2} \textbf{How do algorithm parameters affect surveillance performance?} \\
    Our attack's optimal hash selection outperforms and provides a relative improvement of around 47\% compared to a baseline random hash selection strategy on an average across all location settings. 
    The surveillance rate of our attack increases as the adversary is able to add more poisons varying from 15\% (for 1000 poisons) to 53\% (for 20000 poisons).
    
    \item[\textbf{Q3.}]\label{subsec:q3} \textbf{How do the environmental factors affect the surveillance performance?}\\
    The performance of our attack decreases when there are unseen variations to the scene environmental conditions (e.g., changes in furniture or lighting). However, it is still able to obtain 32\% of images with unseen variations to scene layout and  36\% under unseen lighting conditions (as compared to 57\% under minimal variations).
    
    \item[\textbf{Q4.}]\label{subsec:q4} \textbf{How does the surveillance rate compare with the natural performance of the CSIS system? How can the design of the CSIS system be modified to reduce surveillance risks?}\\
    CSIS systems can detect larger variations of illegal images by increasing the distance threshold for the perceptual hashing function. However, in comparison to the increase in CSIS performance, increasing the threshold causes a larger increase in the surveillance risk. Finally, modifying the current CSIS systems to prevent surveillance would severely limit the system's ability to detect inexact illegal images. For the case of NeuralHash, the CSIS performance would fall from 40\% (current setting) to just 2\%.
    
\end{itemize}

\subsection{Experimental Setup}\label{subsec:expsetup}
A challenge in understanding the surveillance risks in CSIS is determining an appropriate setup without access to illegal content. We believe the following setup helps establish an experimental environment and protocol for future work in this space. We adopt the following settings to evaluate our poisoning attack for physical surveillance.
\subsubsection*{\textbf{Client Side Scanning Parameters.}}
As we do not have access to illicit images, we evaluate our attack using the ImageNet dataset. We use images from the superclass `animal' to approximate the concept of illicit images. A superclass is a broad category in ImageNet which represents multiple classes \cite{tsipras2020imagenet}. Additionally, we use the superclass `artifact' to mimic benign images that a user may upload. From the superclass `animal', we further select two mutually disjoint sets -- (1) to construct the CSIS database $\mathcal{C} (|\mathcal{C}|=100k)$ and (2) to represent the set of illicit images available to the attacker, $X_{inject} (|X_{inject}|=20k)$. The size of $X_{inject}$ cannot be less than the number of poisons since each delivery image needs to be a different illicit image. Therefore, we set the size of $X_{inject}$ to be $20k$ (the maximum poison budget used in our evaluation). We use $10k$ benign images to evaluate the false positive rate of the CSIS system which acts a baseline.

\subsubsection*{\textbf{Scene Image Dataset.}} We evaluate our attack on $6$ physical locations. Four of these locations are popular tourist spots -- The Leaning Tower of Pisa (Italy), The Pyramids at Giza (Egypt), Stone Henge (United Kingdom), and Lennon Wall (Czechia). For each of these locations, we scrape user photographs from Instagram using search queries (\#<location name>). Furthermore, we also perform a manual validation on the collected images to remove any image not captured at the specified location setting. For each location, we split the validated images into two disjoint sets -- (1) Reference Set: scene images available to the adversary, (2) User Set: scene images uploaded by the user. The details for the number of images in each set is shown in Table 1.

We selected the remaining two locations such that we could physically capture photographs of the scene. These were two indoor settings - Room 1 and Room 2. Two of the authors performed data collection from the attacker and the user perspective respectively. For each setting, the attacker captured multiple $6$-minute videos scanning the entire scene. Similarly the user also captured multiple $2$-minute video in the room. We use the frames from these videos to be the Reference and User set respectively. The two authors collected this data using two different phone cameras, since the attacker would not know the camera configuration for all the users. To evaluate for different environmental conditions, we performed this data collection under 3 different lighting settings. We also collected data for 3 different room layouts where we moved the objects and furniture. Finally, we collected an additional user video where the another author was in the field of view to account for any unseen moving objects or people in the user images. Details on the collected data are presented in Table 1. Note that if the attacker is able to physically access the scene, they can capture a significantly larger number of scene images. This improves the attack's surveillance rate as we see later in Section \ref{subsec:a1}.

We ensured that only the authors were present in the recordings of Rooms 1 and 2, and thus, we did not need IRB approval. No other individuals were incidentally recorded in any datasets discussed above. Additionally, only authors played the role of attacker or user.

\begin{table}
    \centering
    \small
    \begin{tabular}{@{}cccc@{}}
    \toprule
    Physical Location & $|\text{Ref.}|/|\text{User}|$ & Ref. Source & User Source \\
    \midrule
    Pisa Tower & $600/150$ &Instagram & Instagram \\
    Pyramids Giza & $700/200$ & Instagram & Instagram \\
    Lennon Wall & $200/100$ & Instagram & Instagram \\
    Stone Henge & $500/150$ & Instagram & Instagram \\
    Room 1 & $45000/10000$ & Sam. S22 & Pixel 7 Pro \\
    Room 2 & $45000/10000$ & iPhone 12 & Sam. Z Flip 4 \\
    \bottomrule
    \end{tabular}
    \caption{Dataset for evaluating physical surveillance poisoning attack.}
    \label{tab:dataset_metrics}
\end{table}

\subsubsection*{\textbf{Perceptual Hash Function.}}
For our CSIS setting, we consider two popular perceptual hash functions - NeuralHash and PDQ. NeuralHash is a deep learning model based on the  MobileNetV3 architecture, which outputs a 128 dimension embedding for an RGB input image. It is trained using contrastive learning such that the embeddings of perceptually similar images (ones that are recolored or scaled versions of each other) are "closer" to each other. The ``closeness'' is generally defined by the euclidean norm or the cosine similarity. The generated embedding is first multiplied with a hashing matrix, and then quantized to output a 96 bit hash. Similar to previous work, we extract the NeuralHash model as well as the hashing matrix from a Mac computer~\cite{breakNeuralHash}. In contrast, PDQ is a DCT based hash function that outputs a 256 bit hash. It applies a series of image transformations followed by quantization. We employ Facebook's official implementation for PDQ~\cite{pdqgithub}.

The original design of Apple CSAM detection does not use a distance function to compares two Neural Hashes; rather, it looks for exact hash collisions. This is because the private set intersection protocol is designed to compute exact intersections. As we are not concerned with the cryptography aspects of CSAM detection in this paper, we adopt a more general approach and use a distance function on NeuralHash outputs, as it is conceivable that future applications of this perceptual hashing algorithm might benefit from inexact matches between a user's image hash and the illegal database of hashes. Section \ref{subsec:a4} indeed shows that using a larger distance threshold can improve detection of syntactically transformed CSAM.

\begin{figure}
    \centering
    \includegraphics[width=0.38\columnwidth]{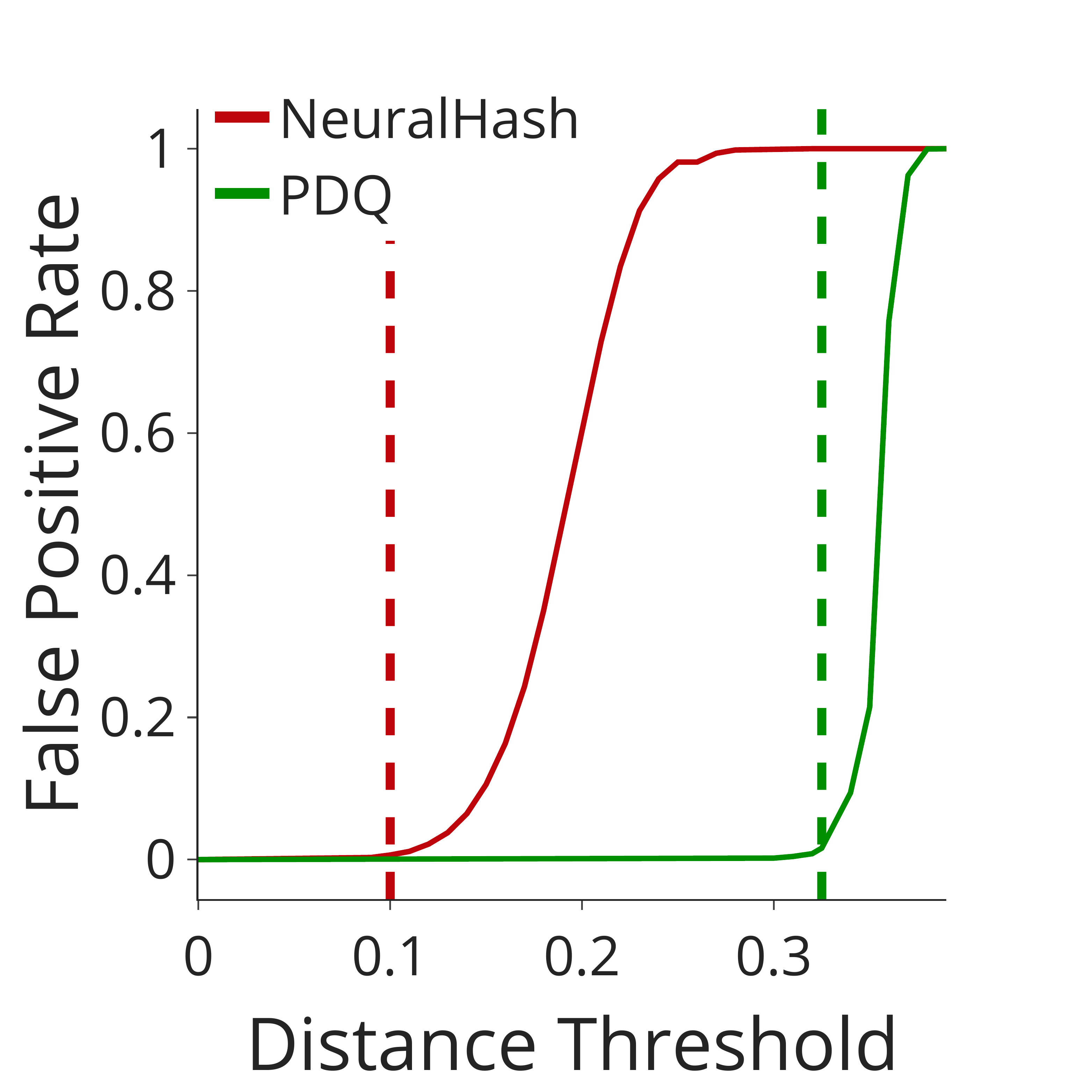}
    \includegraphics[width=0.52\columnwidth]{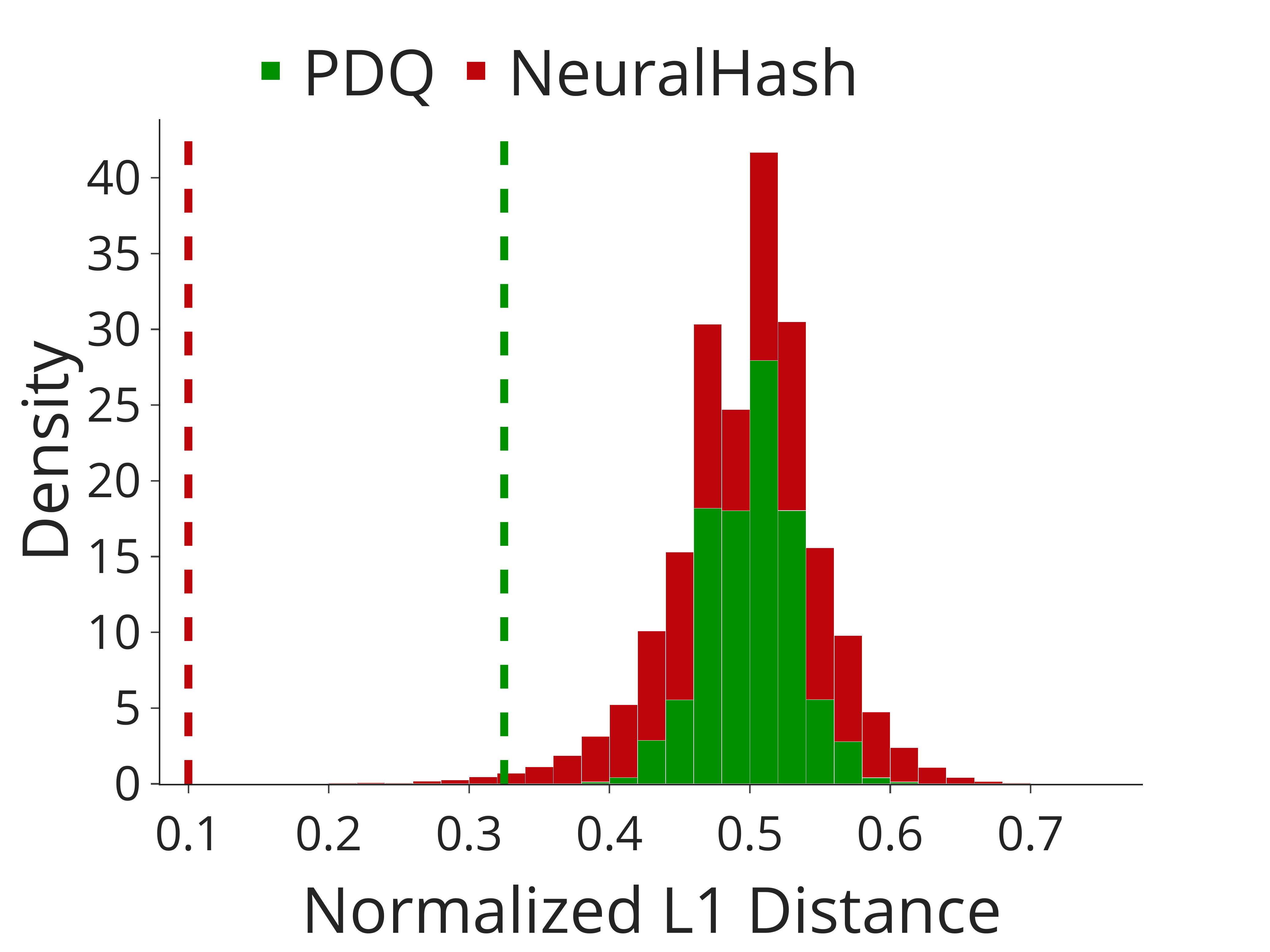}
    \caption{CSIS false positive rate and pairwise hamming distance for the 100k image pairs from the Imagenet  dataset for both NeuralHash and PDQ. We choose the maximum distance thresholds where the False Positive Rate is about to increase (0.1 for NeuralHash and 0.325 for PDQ). The dashed line show the chosen thresholds in both the plots. Note that there is no image pair with pairwise distance less than these threshold values.}
    \label{fig:pairwise}
\end{figure}

To perform more robust detection of illicit images, CSIS would benefit from a larger distance threshold while ensuring that it does not trigger detection on benign images. In Figure \ref{fig:pairwise}, we plot the detection rate of benign images for different distance thresholds. Similar to previous work, we select the inflection point of this curve as our distance threshold. Therefore, for our evaluation, we select a distance threshold of 0.325 for PDQ and 0.1 for NeuralHash (Normalised L1 distances), which incurs a false positive rate of 1\% for both the hash functions. To demonstrate the trade-off between selecting a large threshold and the false positive rate, we select these threshold values such that the false positive rate is slightly above zero. Note that these threshold values are what we selected for our evaluation, the actual thresholds will depend on the desired False Positive Rate of the deployed system.

\subsubsection*{\textbf{Searching for Optimal Hashes.}} Before searching for the optimal poison hashes, we first perform image augmentation for the Reference set of each location. We model physical transformations using random affine transforms and augment the collected scene images to increase the effective size of the Reference set. We perform augmentations such that final number of images is 100000 for each location setting. Next, we search for the optimal poison hashes. We use a poison budget of 5000 (which is 5\% of the CSIS database size) unless stated otherwise. As discussed in Section~\ref{sec:design}, we optimize Equation~\ref{eq:finalopt} using the K-modes clustering algorithm to find the optimal poison hashes \cite{huang1998extensions,cao2009new}. We employ a popular python implementation for K-modes and use the default configuration with 5 random initializations \cite{devos2015}. Furthermore, we compare this with a random selection strategy, where we randomly select scene images hashes to be poisons.

\subsubsection*{\textbf{Generating poison delivery images.}}
For each poison hash, we need to compute the corresponding delivery image by adversarially perturbing a known illicit image such that it's hash is closer to the poison hash. We borrow from previous work on adversarial examples to compute the perturbations. We perform our evaluation for a fixed $L_\infty$ perturbation budget of 8/255. For NeuralHash, we use the white-box PGD attack with 1000 iterations and step size of 0.0001~\cite{croce2020reliable}. Since, PDQ is not differentiable, we use the query-based black-box NES attack with parameters $(\sigma = 0.1, \eta = 0.01)$ and 10000 samples for gradient estimation~\cite{ilyas2018black}. For each poison hash, we attack every image in $X_{inject}$ and choose the perturbed illicit image with the lowest hash hamming distance from the poison hash. We remove the selected image from $X_{inject}$ and continue the attack for the remaining poisons.

\subsubsection*{\textbf{Evaluation Metrics.}} We denote the performance of our approach by the surveillance rate which is the fraction of user images taken at the target surveillance location to be flagged by the CSIS system. Specifically, we use images from the User Set to compute the surveillance rate for each location setting. We also need to ensure that the matching rate for the benign images (i.e., the false positive rate) is low. Furthermore, the CSIS system should still be able to flag illicit images which are syntactically transformed (i.e., natural performance of CSIS). We evaluate the CSIS performance against illicit images under three levels of syntactic variations. Each of these variation levels are represented by a combination of varying degrees of multiple syntactic transformations -- changes in saturation, changes in contrast, changes in brightness and center cropping.

\subsection{Q1. Effectiveness of Physical Surveillance}\label{subsec:a1}
\begin{figure}
    \centering
    \includegraphics[width=\columnwidth]{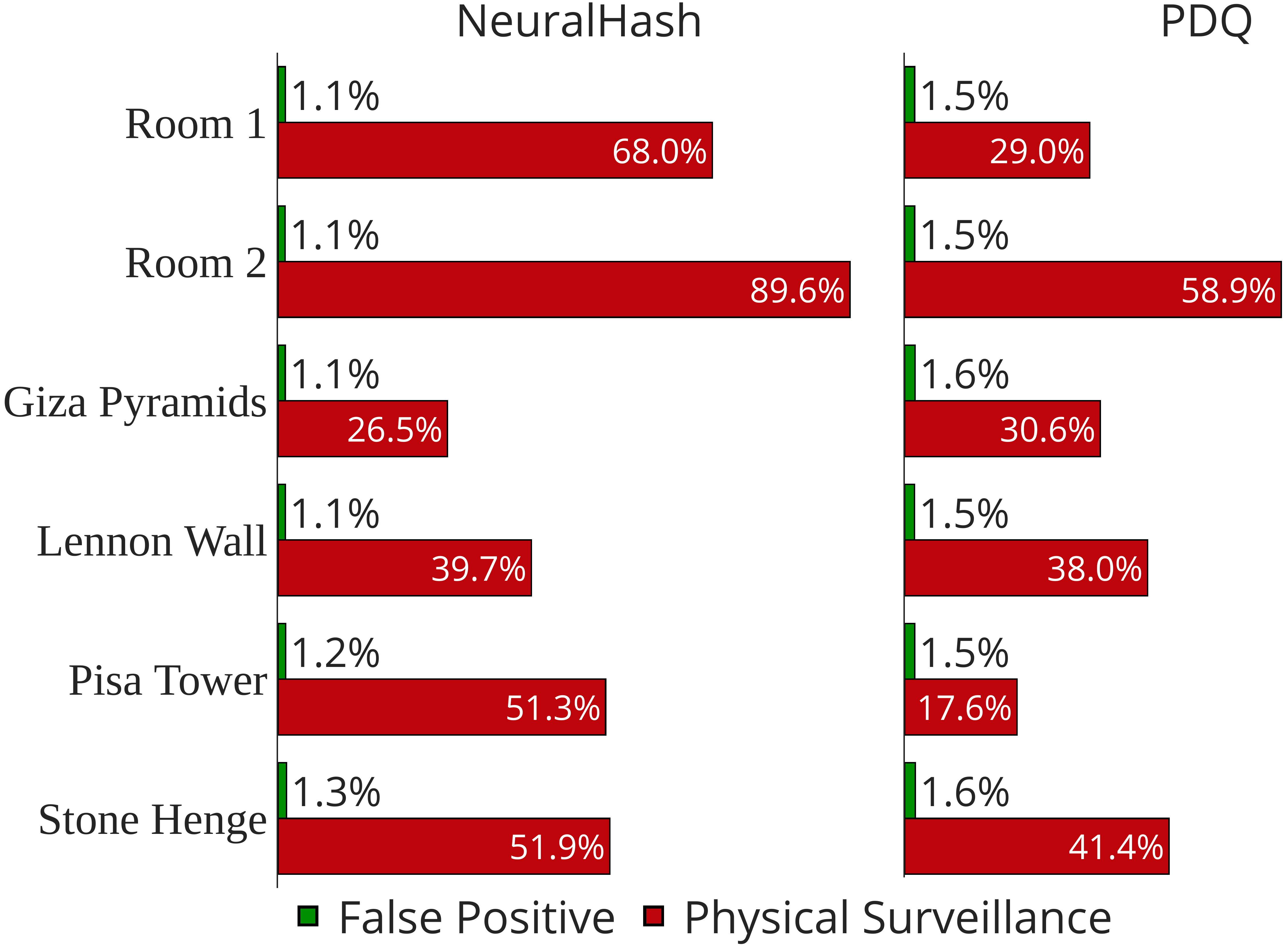}
    \caption{Surveillance and False Positive rates for 4 popular tourist spots as well as 2 physical locations where we physically captured photographs. This shows the effectiveness of the poisoning attack as it achieves high detection rates for targeted locations (Surveillance rate) as compared to the detection of other benign images (False Positive rate).}
    \label{fig:scene_wise}
\end{figure}

We evaluate our approach on real user images from 4 popular tourist spots as well as 2 physical locations where we physically captured photographs. Figure \ref{fig:scene_wise} show the Surveillance rates for all the 6 settings under both NeuralHash and PDQ. Here the distance threshold is set at 0.1 for NeuralHash and 0.325 for PDQ. These threshold values are selected such that the false positive rates are around 1\%. Under all settings, the surveillance rate is significantly higher than false positive rate reaching more than 89\% for the Room 2 NeuralHash setting. Overall, the surveillance rates for NeuralHash are higher as compared to PDQ especially for the Room 1, Room 2 and Pisa Tower settings. Furthermore, the settings where we physically captured photographs (Room 1 and Room 2) observe relatively higher surveillance rates particularly for NeuralHash. This is because the attacker has more control over the reference set images they are collecting. In other experimental settings, the authors were not able to visit those locations to obtain sufficiently detailed reference set images.\footnote{We note that no IRB approval is required for the Room 1 and Room 2 experimental setting because the authors themselves played the roles of user and attacker.}  In Table \ref{tab:attack_instance}, we show sample images for an attack instance on a NeuralHaash based CSIS system for 2 location settings -- Room 1 and Room 2. In summary, this experiment demonstrates that an attacker can conduct physical surveillance in a way that approximates them placing a camera at the target location.

\begin{table*}
    \centering
    \small
    \begin{tabular}{@{}c|c|c|c@{}}
    \toprule
    \textbf{Location Setting} & \textbf{Attacker Profiled Images} & \textbf{Poison Delivery Images} & \textbf{Detected Images} \\
    \midrule
    Room 1 & \includegraphics[width=0.1\linewidth]{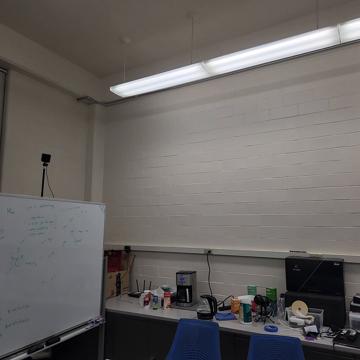}\;\includegraphics[width=0.1\linewidth]{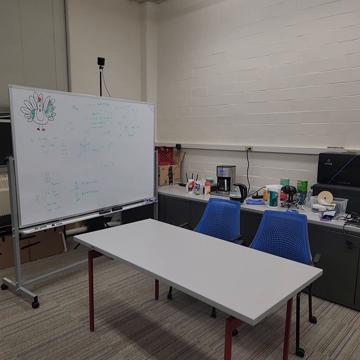}\;\includegraphics[width=0.1\linewidth]{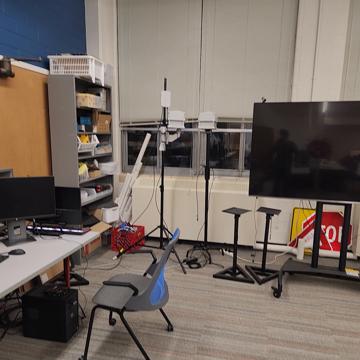}
     & \includegraphics[width=0.1\linewidth]{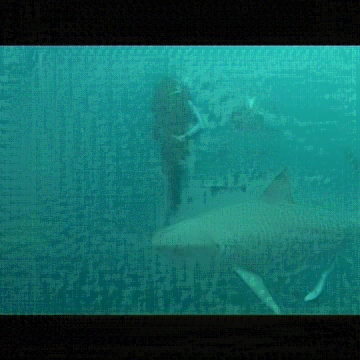}\;\includegraphics[width=0.1\linewidth]{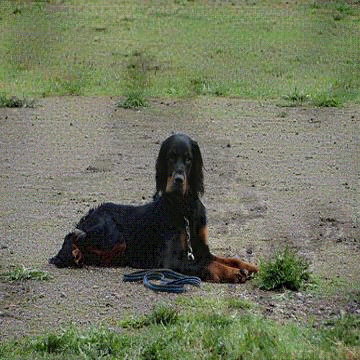} & \includegraphics[width=0.1\linewidth]{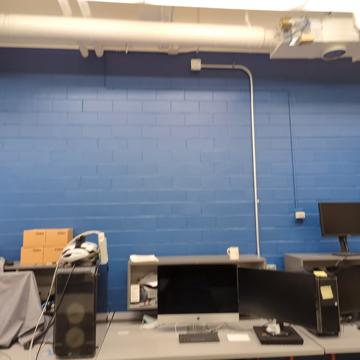}\;\includegraphics[width=0.1\linewidth]{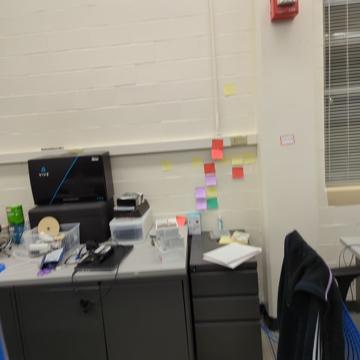}\\
     \midrule
    Room 2 & \includegraphics[width=0.1\linewidth]{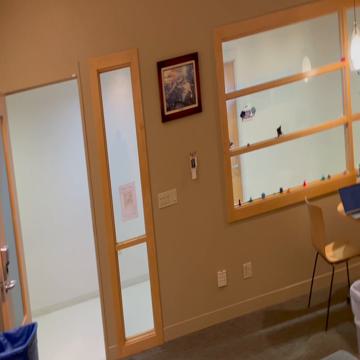}\;\includegraphics[width=0.1\linewidth]{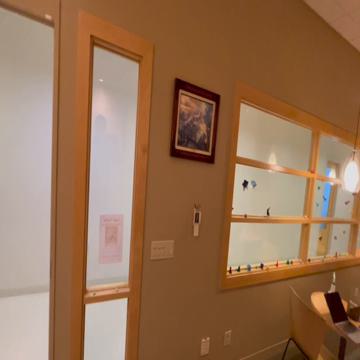}\;\includegraphics[width=0.1\linewidth]{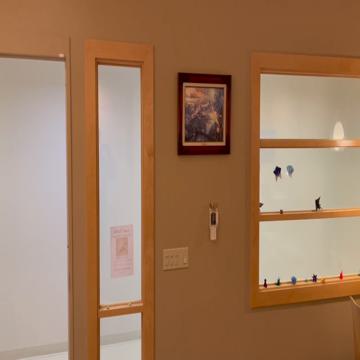}
     & \includegraphics[width=0.1\linewidth]{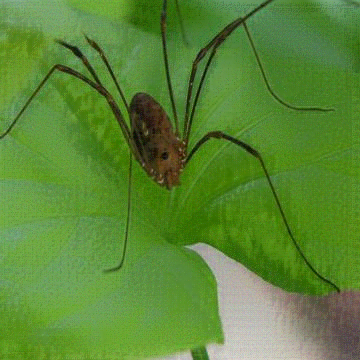}\;\includegraphics[width=0.1\linewidth]{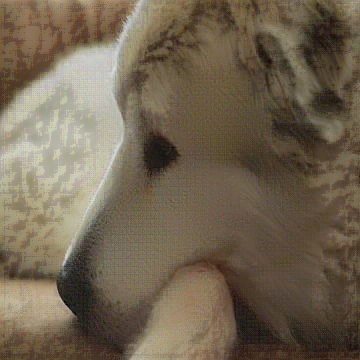} & \includegraphics[width=0.1\linewidth]{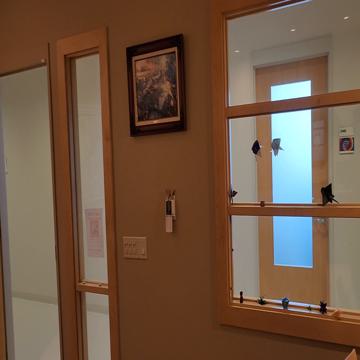}\;\includegraphics[width=0.1\linewidth]{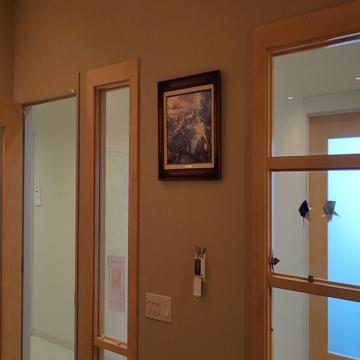}\\
    \bottomrule
    \end{tabular}
    \caption{Sample images for surveillance attack on a CSIS system based on NeuralHash for location settings --- Room 1 and Room 2. The poison delivery images have been generated using white-box PGD-1000 attack with a $L_{\infty}$ perturbation budget of $8/255$.}
    \label{tab:attack_instance}
\end{table*}


\subsection{Q2. Effect of Algorithm Parameters on Surveillance Success}\label{subsec:a2}
We also evaluate how the different attack parameters affect the surveillance rate. Table \ref{fig:poison_table} shows the surveillance rates for each location setting for multiple poison budgets. We also compare our optimal hash selection strategy of using K-modes clustering against a baseline strategy where the hashes are randomly selected from the hashes of the augmented scene images. We make the following observations. First, the surveillance rate significantly increases with increasing number of poisons for both PDQ and NeuralHash. Second, the rate of increase is not uniform across the different locations. The increase is particularly significant for the settings Room 1 and Room 2. This again points to the fact that the adversary is more powerful if they can physically capture photographs from the scene. We can also observe that the surveillance rate is significantly higher for the k-modes optimal hash selection strategy as compared to random selection providing a relative improvement of around 47\% on average across all the location settings. The relative increase is particularly higher when the poison budget is lower. For instance, for Room 1 the surveillance rate between the two strategies differs by almost 50\% for a poison budget of 1000. However, the difference reduces to only around 15\% for a budget of 20000. This is because compared to the random strategy, K-modes is most effective when the size of the poison budget is relatively smaller as compared to the size of the Augmented Reference Set (size=100000). In this case, a poison budget of 100000 would select the whole Augmented Reference Set for both strategies.

\begin{table*}[t]
\small
\begin{center}\scriptsize
\begin{tabular}{lp{0.5in}cccc|cccc}
\toprule
Scene                                  & Poison Strategy & \multicolumn{8}{c}{Number of Poisons} \\ 
& & \multicolumn{4}{c}{PDQ}  & \multicolumn{4}{c}{NeuralHash} \\
                                       &        & 1000 (1\%)   & 5000 (5\%) & 10000 (10\%) & 20000 (20\%) &    1000 (1\%)   & 5000 (5\%) & 10000 (10\%) & 20000 (20\%)\\ \hline

\multirow{2}{0.5in}{Pisa Tower} & Random          & $0.04 \pm 0.02$     & $0.09 \pm 0.03$  & $0.13 \pm 0.00$  & $0.17 \pm 0.01$  & $0.07 \pm 0.01$  
& $0.24 \pm 0.01$     & $0.34 \pm 0.02$  & $0.47 \pm 0.01$\\
                                       & K-Modes         & $0.11 \pm 0.01$     & $0.18 \pm 0.00$  & $0.20 \pm 0.01$  & $0.21 \pm 0.01$  & $0.22 \pm 0.01$  
& $0.40 \pm 0.01$     & $0.45 \pm 0.03$  & $0.52 \pm 0.01$\\ \hline
\multirow{2}{0.5in}{Lennon Wall}           & Random          & $0.12 \pm 0.02$     & $0.31 \pm 0.06$  & $0.40 \pm 0.04$  & $0.46 \pm 0.01$  & $0.11 \pm 0.01$  
& $0.22 \pm 0.02$     & $0.28 \pm 0.03$  & $0.34 \pm 0.02$\\
                                       & K-Modes         & $0.24 \pm 0.02$     & $0.38 \pm 0.02$  & $0.42 \pm 0.01$  & $0.44 \pm 0.01$  & $0.19 \pm 0.01$  
& $0.28 \pm 0.03$     & $0.34 \pm 0.01$  & $0.39 \pm 0.01$\\ \hline
\multirow{2}{0.5in}{Stone Henge}           & Random          & $0.14 \pm 0.03$     & $0.33 \pm 0.04$  & $0.45 \pm 0.02$  & $0.55 \pm 0.01$  & $0.09 \pm 0.03$  
& $0.26 \pm 0.04$     & $0.38 \pm 0.05$  & $0.50 \pm 0.02$\\
                                       & K-Modes         & $0.23 \pm 0.02$     & $0.41 \pm 0.01$  & $0.46 \pm 0.02$  & $0.51 \pm 0.01$  & $0.20 \pm 0.01$  
& $0.37 \pm 0.01$     & $0.46 \pm 0.01$  & $0.53 \pm 0.01$\\ \hline
\multirow{2}{0.5in}{Giza Pyramids}         & Random          & $0.12 \pm 0.02$     & $0.22 \pm 0.02$  & $0.28 \pm 0.03$  & $0.32 \pm 0.03$  & $0.05 \pm 0.01$  
& $0.11 \pm 0.01$     & $0.19 \pm 0.03$  & $0.23 \pm 0.02$\\
                                       & K-Modes         & $0.22 \pm 0.01$     & $0.31 \pm 0.01$  & $0.33 \pm 0.00$  & $0.35 \pm 0.01$  & $0.15 \pm 0.01$  
& $0.23 \pm 0.01$     & $0.26 \pm 0.02$  & $0.29 \pm 0.01$\\ \hline
\multirow{2}{0.5in}{Room 1}           & Random          & $0.05 \pm 0.01$     & $0.21 \pm 0.03$  & $0.32 \pm 0.04$  & $0.46 \pm 0.04$  & $0.15 \pm 0.01$  
& $0.40 \pm 0.03$     & $0.53 \pm 0.03$  & $0.63 \pm 0.04$\\
                                       & K-Modes         & $0.08 \pm 0.02$     & $0.29 \pm 0.08$  & $0.41 \pm 0.11$  & $0.51 \pm 0.12$  & $0.29 \pm 0.05$  
& $0.57 \pm 0.06$     & $0.66 \pm 0.07$  & $0.73 \pm 0.06$\\ \hline
\multirow{2}{0.5in}{Room 2}           & Random          & $0.16 \pm 0.01$     & $0.45 \pm 0.11$  & $0.54 \pm 0.12$  & $0.63 \pm 0.10$  & $0.59 \pm 0.09$  
& $0.77 \pm 0.06$     & $0.80 \pm 0.05$  & $0.84 \pm 0.03$\\
                                       & K-Modes         & $0.32 \pm 0.05$     & $0.59 \pm 0.12$  & $0.68 \pm 0.13$  & $0.72 \pm 0.12$  & $0.74 \pm 0.03$  
& $0.85 \pm 0.02$     & $0.88 \pm 0.03$  & $0.90 \pm 0.02$\\ \hline
\end{tabular}
\end{center}
\caption{Comparing surveillance rates against a baseline random poison selection strategy for different poison budgets. K-Modes strategy for poison selection outperforms the random baseline for all settings. Increasing the poison budget significantly improved the surveillance rate specially for the Room 1 and Room 2 settings. }
\label{fig:poison_table}
\end{table*}

\begin{figure}
    \centering
    \includegraphics[width=\linewidth]{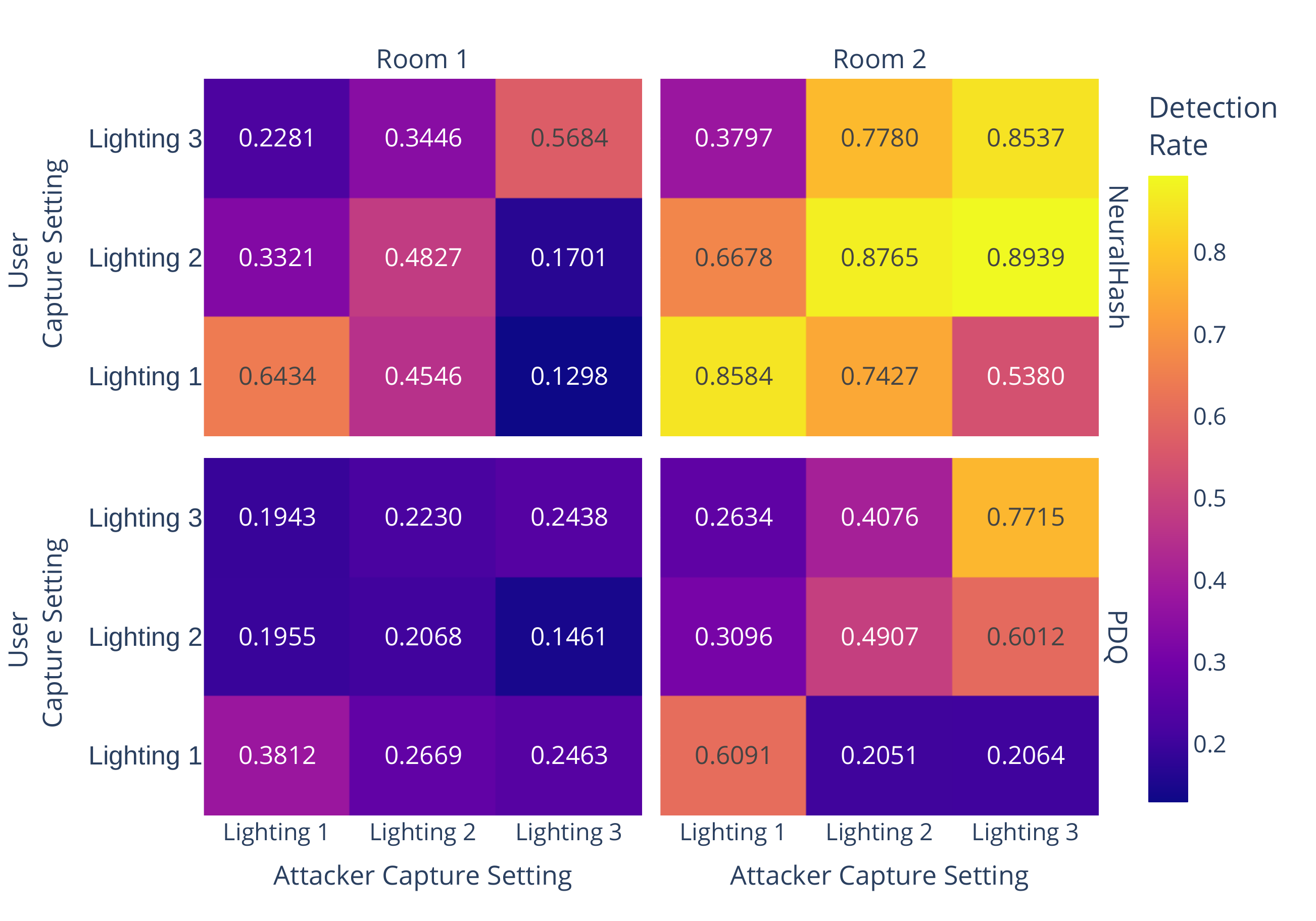}
    \caption{Surveillance rates for cross evaluation of 3 different lighting conditions. Evaluating on a different lighting condition reduces surveillance rate but it is still $> 14\%$ which is much higher than the false positive rate of $~1\%$. Moreover, surveillance on NeuralHash based CSIS systems is more robust to lighting changes as compared to PDQ.}
    \label{fig:light_confusion}
\end{figure}

\begin{figure}
    \centering
    \includegraphics[width=\linewidth]{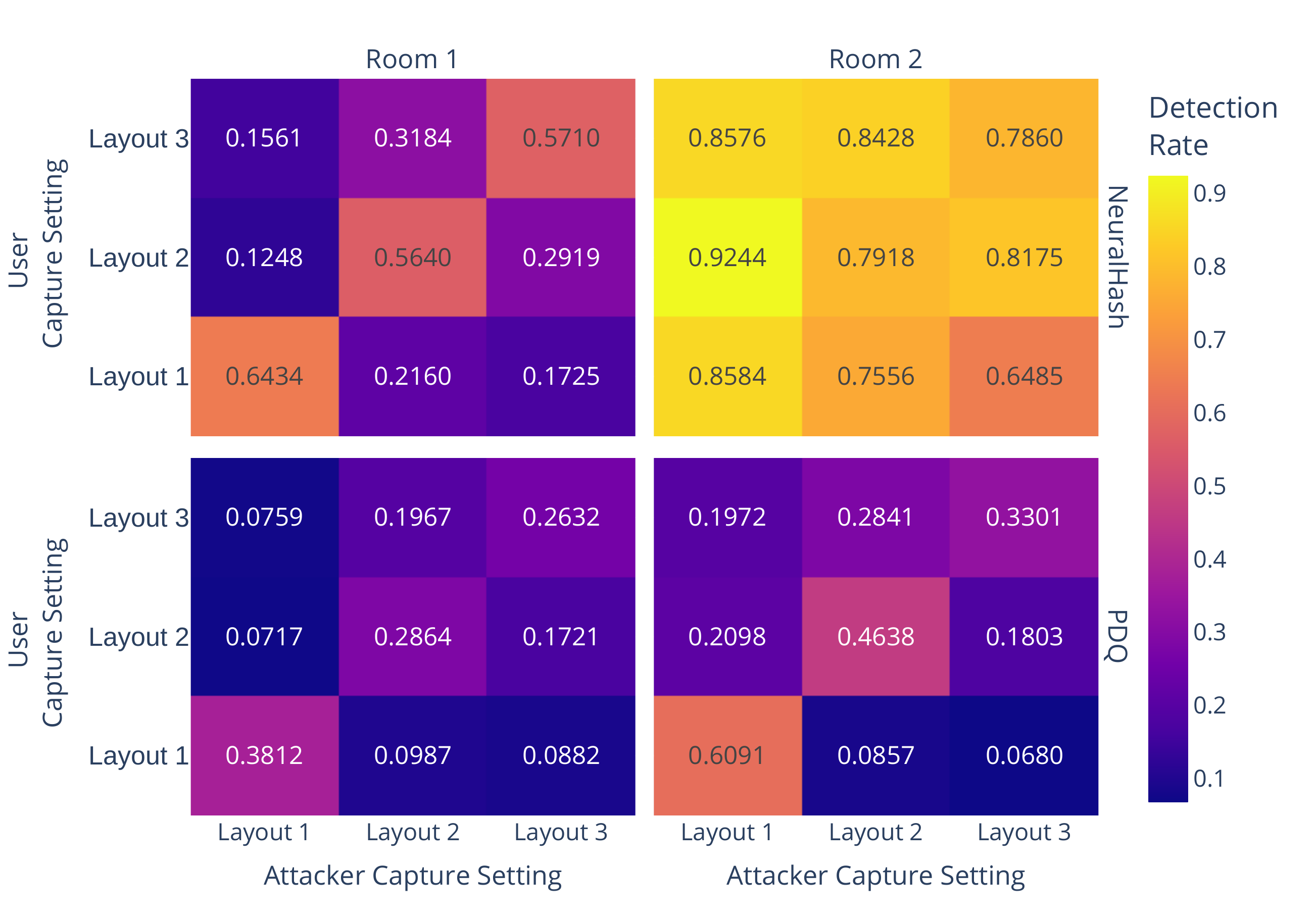}
    \caption{Surveillance rates for cross evaluation of 3 different layout conditions. Evaluating on a different layout condition leads to lower surveillance rates as compared to those for different lighting setting.  Moreover, surveillance on NeuralHash based CSIS systems is more robust to layout changes as compared to PDQ.}
    \label{fig:layout_confusion}
\end{figure}

\subsection{Q3. Environmental Factors}\label{subsec:a3}
Next, we evaluate how changes in the environmental conditions such as lighting (Figure~\ref{fig:light_confusion}) and scene layout (Figure~\ref{fig:layout_confusion}) affect the surveillance rate. We perform an ablation study by cross evaluating three different lighting and scene layout settings. We generate the poisons using the Reference Set from each lighting setting and evaluate the performance against the User Set of each of the 3 lighting settings. This results in a total of 9 experimental instances. A similar analysis is performed for each of the layout settings. First, for both lighting and layout, the surveillance rate is the highest if the Reference Set and the User Set belong to the same lighting or layout setting. By contrast, when the Reference set and User set belong to different lighting or layout settings, there is a decrease in surveillance rate. This experiment evaluates the scenario where the user uploads images in environment conditions that were not accounted for when the attacker scanned the scene. 
NeuralHash incurs a relative decrease in surveillance of 34\% for unseen lighting and 27\% for unseen layout. In contrast, PDQ suffers a relative decrease in surveillance of 40\% for unseen lighting and 65\% for unseen layout. This suggests that surveillance is more stable under unseen environment conditions for NeuralHash as compared to PDQ. Additionally, we can observe that the detection performance is slightly more robust to lighting changes as compared to layout changes. This is likely due to the underlying hash being more robust to brightness changes as compared to translations and rotations.

\begin{figure}
    \centering
    \includegraphics[width=\linewidth]{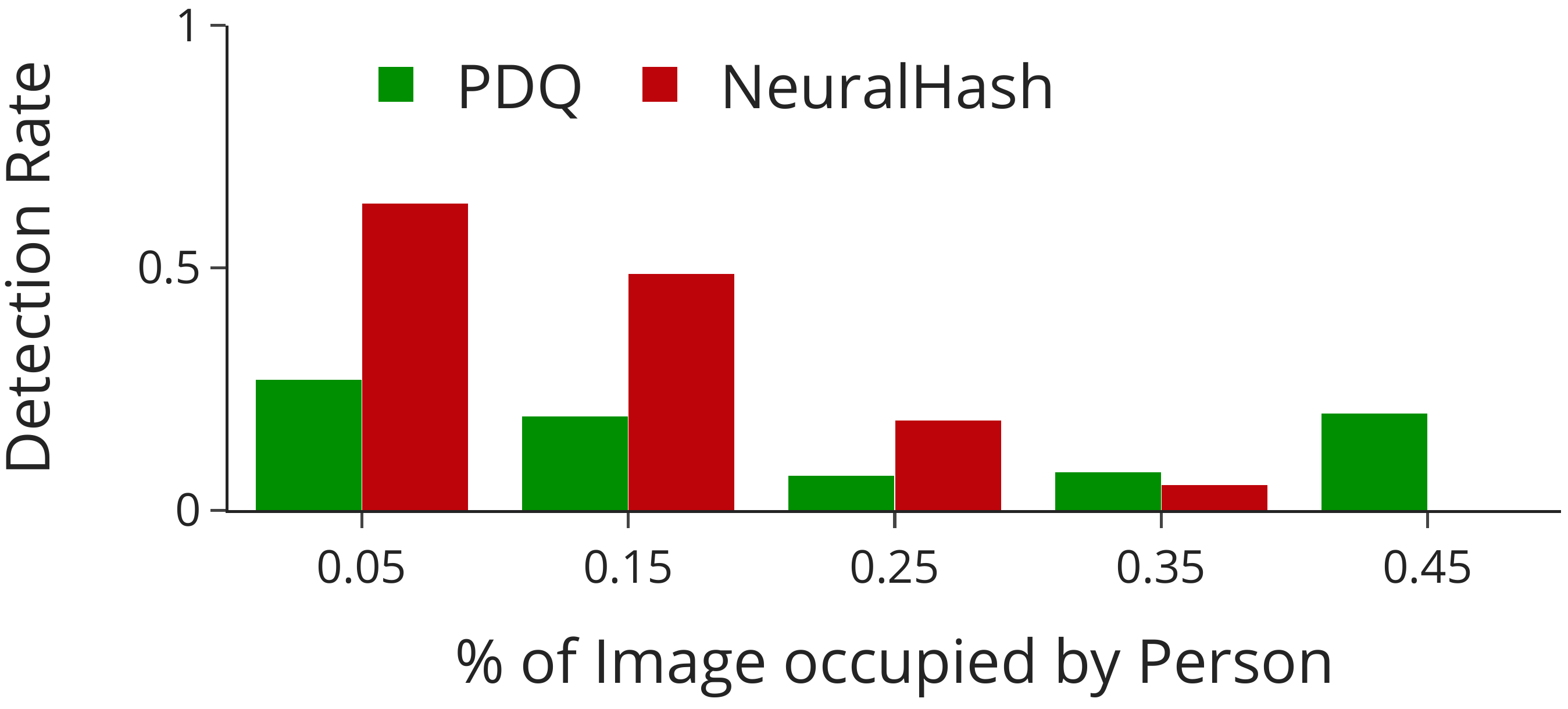}
    \caption{Surveillance rates of image frames with varying amount of field of view occupied by a person. Note that here the poisons have been generated using scene images without any person. The surveillance rate gradually decreases as more area of the background get obfuscated, but the attack still provides surveillance rate of more than $\>20\%$ even when more than a quarter of the frame is occupied by a person.}
    \label{fig:person_plot}
\end{figure}

Next, we evaluate how the attack performance is affected when a unseen person is present in the captured photo. Note that the first 4 experiment settings where images are scraped from Instagram already include people in the Field of View (FoV) of various proportions, showing that indeed, surveillance is possible with people in the image. In this experiment, we isolate the effect of a person's presence and study the effect on surveillance rate while controlling the percentage of the FoV that is occupied by a person. Moreover, for this experiment, the poison hashes have been generated with a scan of only the background and no person in any of the collected images. The results in Figure \ref{fig:person_plot} show that the surveillance rate gradually decreases as the FoV of the person in the foreground increases. Specifically, the surveillance rate for NeuralHash decreases from 65\% to 22\% as the FoV increases from 0 to a quarter of the image. Subsequently, it goes down to 0 as the FoV increases beyond 35\%. Similar trend hold for PDQ. This experiment highlights the risks associated with physical surveillance as the detected images leak privacy of the persons captured in the photo.

\begin{figure*}
    \centering
    \subfigure[NeuralHash]{
    \includegraphics[width=0.47\linewidth]{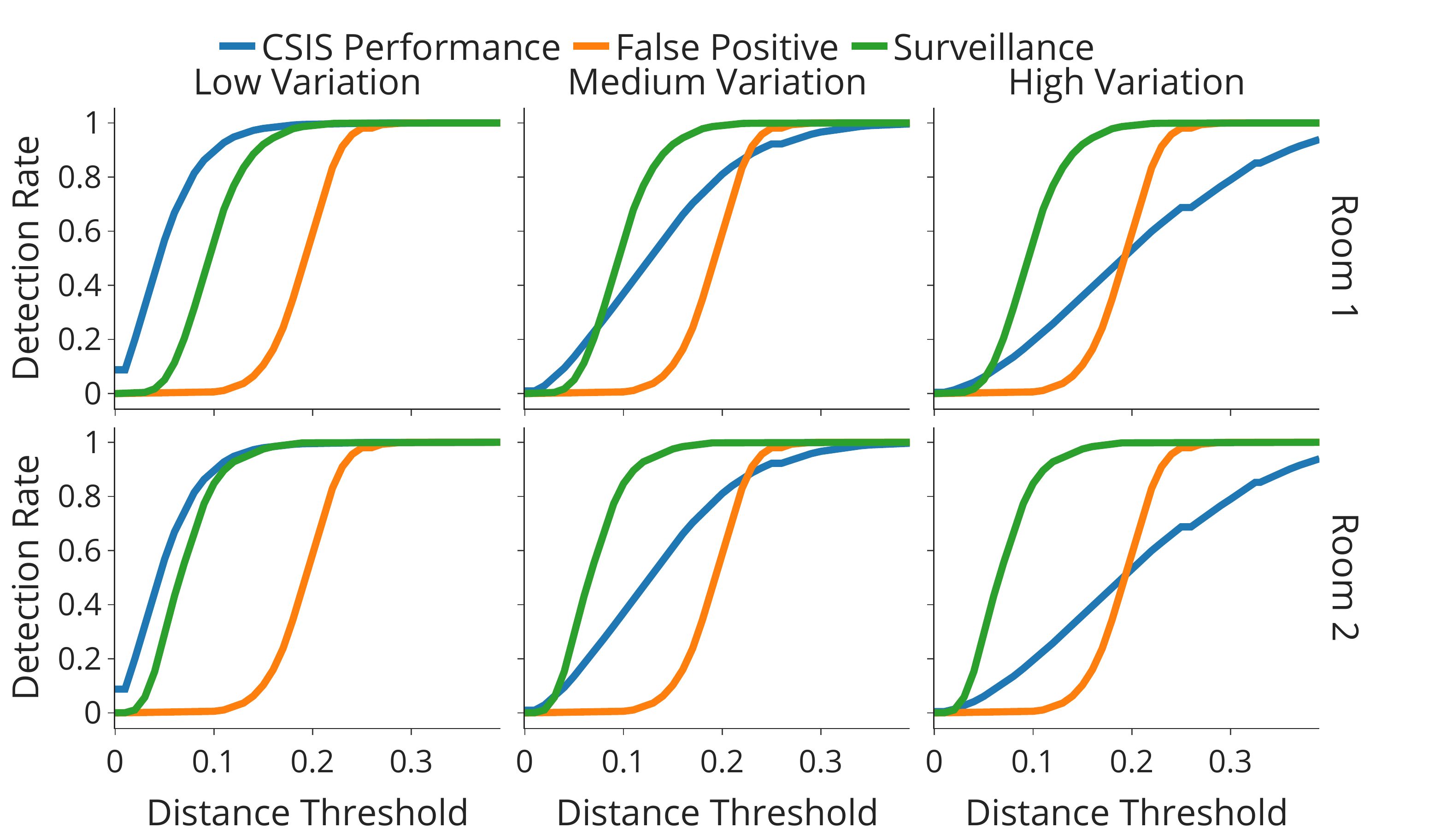}}
    \subfigure[PDQ]{
    \includegraphics[width=0.47\linewidth]{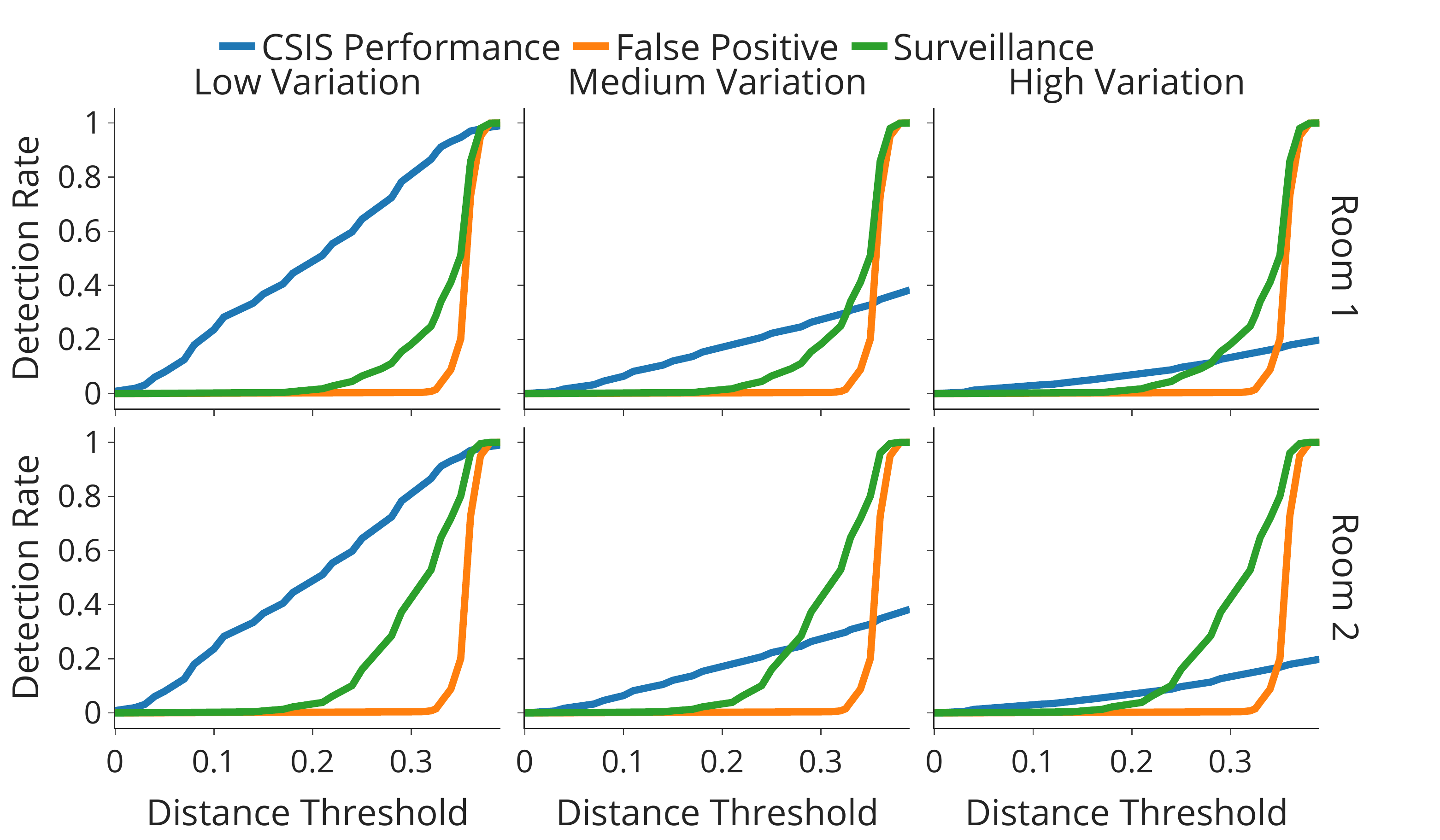}}
    \caption{Illegal Image Detection (CSIS Performance), False Positive Rate and Surveillance rates for varying distance thresholds under 3 different image variation settings. We observe that the rate of increase of surveillance rate is higher than that of CSIS performance.}
    \label{fig:trade-off}
\end{figure*}

\begin{table}[H]
    \centering
    \small
    \begin{tabular}{@{}cc|c|c@{}}
    \toprule
    Transformation & Low & Medium & High \\
    \midrule
    Brightness & 0.9/1.1 & 0.7/1.3 & 0.5/1.5\\
    Contrast & 0.9/1.1 & 0.7/1.3 & 0.5/1.5\\
    Saturation & 0.9/1.1 & 0.7/1.3 & 0.5/1.5\\
    Center Crop & 0.9/1.0 & 0.7/1.0 & 0.5/1.0\\
    \bottomrule
    \end{tabular}
    \caption{Syntactic transformations to evaluate CSIS performance.}
    \label{tab:syntactic_transformation}
\end{table}

\subsection{Q4. Trade-off between CSIS Robustness and Surveillance}\label{subsec:a4}

Our goal in this experiment is to analyze the trade-off between how well a CSIS system detects illegal images and how well a surveillance adversary can achieve their goals. To do this, we examine detection rate, false positive rate and surveillance rate by varying the distance threshold, because this controls the natural performance of the CSIS system --- higher distance thresholds give the system more invariance to syntactic transformations of the illegal material making it harder for adversaries to evade the system. For each threshold, we analyze the CSIS performance against image variations under which a robust CSIS system must operate. For this, we consider three different classes of syntactic transformations  (Table~\ref{tab:syntactic_transformation} shows the parameter ranges for the image transformations used in the experiment, such as brightness, contrast and saturation). Figure~\ref{fig:trade-off} documents the results of this experiment. We observe that both surveillance rate and CSIS performance increase with increasing distance threshold but the slope of surveillance rate is higher specially for medium and high variation settings. 

Next, we look at how this analysis impacts the design decision of the CSIS system. Without considering the risk of surveillance, a CSIS system is designed to maximize the performance of illegal image detection while incurring a tolerable false positive rate (this is also described in Section \ref{subsec:expsetup}). To do this, we can choose the largest distance that allows for a tolerable false positive rate. For NeuralHash it would be 0.1, which achieves a CSIS performance of around 90\%, 40\% and 20\% for the low, medium and high variation settings. For PDQ, the desired threshold would be around 0.32 achieving CSIS performance of around 80\%, 30\% and 15\%. However, we show in previous sections that these threshold values pose a high surveillance risk of >40\% for both NeuralHash and PDQ. To defend against surveillance attacks, a CSIS system now needs to be designed with an additional objective of reducing the surveillance rate. To do so, we need to choose a much lower distance threshold of around 0.02 for NeuralHash and 0.17 for PDQ. However, this significantly reduces the CSIS performance. For NeuralHash, it reduces to around 10\%, 2\%, 0\% for the three variation settings, whereas for PDQ, it reduces to 35\%, 15\% and 5\%. This means that preventing surveillance attacks would lead to a significant reduction in the performance of the CSIS system.

\section{Related work}

\subsection{Client-Side Image Scanning and its Risks}

Motivated by the impending roll out of end-to-end encryption for services like iCloud~\cite{icloud-encryption}, there are several proposals to adapt server-side illegal content scanning technologies to work directly on the client-side~\cite{EFFApple}. These systems side-step the issue of end-to-end encryption by scanning user photos before it is encrypted and sent to the cloud. These systems essentially work a backdoor into the encryption scheme that allows selective opening of encryptions based on whether the underlying content matches known illegal content. For example, Apple NeuralHash is the most prominent example of a CSIS that uses a perceptual hashing function and a database of illegal image hashes to determine whether an encrypted image can be decrypted by the service provider. Despite their potential benefits in curbing the distribution of illegal content, they have faced criticisms~\cite{bugsinpocket}.

One class of criticism exploits the fact that perceptual hash functions are not robust to adversaries~\cite{jain2022adversarial,breakNeuralHash}. For example, it is straightforward to modify illegal images such that their corresponding hashes do not match with the illegal image database, allowing criminals to easily continue the distribution of such content~\cite{breakNeuralHash}. Another class shows that the converse is possible --- benign images can be subtly altered so that their corresponding perceptual hash matches an entry in the database, leading to its eventual decryption~\cite{cryptoeprint:2021/1531}. The attacks represent a serious loss of privacy and a total defeat of the goals of end-to-end encryption.

Our work contributes to the conversation around the risks of CSIS technology. Specifically, we identify a new type of physical surveillance attack that allows malicious actors to monitor physical locations by tapping into the photos that unwitting users take at those locations. Our work takes inspiration from the existing yet hypothetical criticisms of CSIS and provides experimental evidence of the extent to which such surveillance is possible. 

Prokos et al. recently outlined a taxonomy of surveillance threats on CSIS technology~\cite{cryptoeprint:2021/1531}. Specifically, they show how an adversary can detect the presence of a specific surveillance image that is legal content but of interest to the adversary using a collision attack on the perceptual hash function. Translated to our setting, this image could be a photograph that the user takes at a target location. However, the crucial difference is that we are interested in detecting a distribution of images taken at a specific location, not just a single specific image. Therefore, we expand the taxonomy of Prokos et al. by contributing a new type of physical surveillance attack by poisoning the illegal image hash database of a CSIS system using a novel k-modes formulation.

\subsection{Methods of finding Perceptual Hashing collisions}

Techniques to find colliding images vary depending on the implementation of the perceptual hashing function. Prokos et al. use a gradient approximation to find targeted second pre-image attacks on PhotoDNA and PDQ~\cite{cryptoeprint:2021/1531}.
Qingying et al. apply blackbox methods to craft evasion attacks on perceptual hashing algorithms pHash and Blockhash ~\cite{hao2021s}.
NeuralHash is a variant of the MobileNetV3 neural network, so the attacks use common gradient-based methods and more advanced generative adversarial network based preimage attacks~\cite{breakNeuralHash}. Our work uses these attack algorithms to compute poison delivery images. However, we note that our k-modes algorithm to compute optimal poison hashes does not rely on how the perceptual hash function works internally.

\section{Discussion}

\noindent\textbf{Defenses against physical surveillance.} We adopt a systems-view of the problem and reason about how various stages of the CSIS pipeline can work together to make physical surveillance attacks harder. First, one could leverage the recent progress in defenses against adversarial examples to make computing poison delivery images harder. For example, techniques like adversarial training~\cite{madry2017towards}, diffusion-based adversarial purification~\cite{nie2022DiffPure} or certified robustness~\cite{certified-defenses-aditi,lecuyer2019certified} can increase the distortion required on the adversarial example to the point that either the human curator rejects the sample as being too noisy or the resultant hash of the poison delivery image is too far from the desired hash. The challenge is that such techniques would work for deep learning-based perceptual hashes like NeuralHash but not for algorithms like PDQ. 

Second, we could augment the CSIS pipeline with an out-of-distribution (OOD) detector~\cite{fang2022learnable,cai2023frequency,sun2022dice}. An OOD algorithm learns to detect data that falls outside a specific distribution. In our case, illegal material is considered to be in-distribution and anything else is out-of-distribution. This increases the bar on the attacker in that they have to now cause perceptual hash collisions and simultaneously trick the OOD algorithm into believing the adversarial image is within the expected distribution. 

\noindent\textbf{Targeting multiple locations for surveillance.} Our experiments focus on a single target location at any given point in time. It is possible for an attacker to simultaneously conduct surveillance on multiple locations. They would have to run our k-modes and poison delivery image algorithms for each location to compute the set of poisons and then submit them to the curator. This will increase the fraction of the hash database that is poisoned and it can make curators suspicious if a single attacker is submitting a very large number of illegal images. Over time, this can make the illegal hash database grow without bounds as there is no way to request deletion of hashes from the database currently. 

\noindent\textbf{A physical object that can cause collisions.} In adversarial machine learning, there are physical objects that can trick classifiers and object detectors into outputting attacker-desired decisions, such as a `Stop' sign with stickers that forces a classifier to output `Speed Limit'~\cite{https://doi.org/10.48550/arxiv.1707.08945}. One could hypothesize the existence of a similar object for perceptual hashing such that the resultant hash of the scene would always collide with a fixed pre-determined value. This can make the surveillance attack easier --- the attacker has to fabricate an object and simply place it in the target location and only poison the hash database with a single value. This object could be as simple as an abstract piece of art. We have attempted to fabricate such an object by borrowing techniques from physical adversarial examples, but have not been successful. We suspect that this is because perceptual hashing functions like NeuralHash and PDQ are only invariant to small syntactic transformations and cannot generalize to larger semantic features of a scene. In contrast, machine learning image classifiers are trained to extract high level features of an image making them highly invariant to semantic transformations, which leads to the possibility of physical adversarial examples. A recent work which compares the success of expectation over transformation (the primary technique used to generate physical adversarial examples) with the semantic invariance of the model, corroborates the concept~\cite{gao2022on}.

\section{Conclusion}

Client-side image scanning systems aim to curb the distribution of illegal content but face strong criticism about the risks associated with the technology. We have provided experimental evidence for one specific type of risk --- malicious nation states or law enforcement agencies can misuse CSIS to conduct a form of physical surveillance. Specifically, they can obtain a surveillance capability the approximates the idea of placing a camera in a target location. We show a simple hash database poisoning algorithm that can achieve surveillance rates upwards of 40\% with approx. 5\% of the hash database being poisoned. We also characterize a tension between the robustness of CSIS performance and surveillance success rate --- if a designer wishes to make a CSIS system less vulnerable to physical surveillance, it is likely that the performance of CSIS on actually detecting illegal content will decrease. This suggests an undesirable trade-off --- scan robustly for illegal content while being vulnerable to physical surveillance.  We do not take a specific stance on whether the existence of such physical surveillance attacks imply that CSIS technology should not be deployed. Rather, our goal is to inform the public debate around the use of such technology and balance it with experimental evidence of one type of risk. 

{\small
\bibliographystyle{ieee_fullname}
\bibliography{main}
}

\end{document}